\documentclass[twocolumn,pre]{revtex4}
\usepackage{epsfig,graphicx}
\usepackage[all]{xy}
\newcommand{\ep}{\varepsilon}
\begin{document}
\title{Multi-Scaling of Correlation Functions in Single Species Reaction-Diffusion Systems}
\author{Ranjiva M. Munasinghe} 
\affiliation{Department of Mathematics, University of Warwick, Gibbet Hill Road, Coventry, CV4 7AL, UK} 
\email{ranm@maths.warwick.ac.uk}
\author{R. Rajesh}
\affiliation{Martin Fisher School of Physics, Brandeis University}
\email{rrajesh@brandeis.edu}
\author{Oleg V. Zaboronski}
\affiliation{Department of Mathematics, University of Warwick}
\email{olegz@maths.warwick.ac.uk}

\begin{abstract}
We derive the multi-fractal scaling of probability distributions of multi-particle configurations for the 
binary reaction-diffusion system $A+A \rightarrow \emptyset$ in $d \leq 2$ and
for the ternary system $3A \rightarrow \emptyset$ in $d=1$. For the binary reaction we find that 
the probability $P_{t}(N, \Delta V)$ of finding $N$ particles 
in a fixed volume element $\Delta V$ at time $t$ decays in the limit of large time as $(\frac{\ln t}{t})^{N}(\ln t)^{-\frac{N(N-1)}{2}}$ for $d=2$ and
$t^{-Nd/2}t^{-\frac{N(N-1)\varepsilon}{4}+\mathcal{O}(\ep^2)}$ for $d<2$. Here $\ep=2-d$. For the ternary reaction in one dimension we find that $P_{t}(N,\Delta V) \sim
(\frac{\ln t}{t})^{N/2}(\ln t)^{-\frac{N(N-1)(N-2)}{6}}$. 
The principal tool of our study is the dynamical renormalization group. We compare predictions of $\ep$-expansions for $P_{t}(N,\Delta V)$ for 
binary reaction in one dimension
against exact known results. We conclude that the $\ep$-corrections of order two and higher are absent in the above answer for $P_{t}(N, \Delta V)$ for $N=1,2,3,4$. 
Furthermore we conjecture the absence of $\ep^2$-corrections for all values of $N$.   
\end{abstract}

\maketitle

\section{Introduction}

Modeling chemical reactions is an important practical and theoretical problem. Systems of reacting particles are typical
of complex irreversible non-equilibrium systems. The popular description of these systems in terms of simple rate equations
currently adopted in chemical kinetics \cite{house} does not appear to have a firm theoretic foundation and often produces wrong results.

Interacting particle systems provide a good model for simple chemical reactions. Well-known examples include systems of diffusing-coalescing
and diffusing-annihilating particles describing reactions
$A+A \rightarrow A$ and $A+A \rightarrow \emptyset$ respectively. Extensive studies of these particle systems in low dimensions have shown that 
the rate equations yield incorrect results
for the computation of average concentrations of reactants, see \cite{priv} for review. Rate equations fail in low dimensions due to the presence
of large fluctuation effects, which violate mean-field theory (MFT) assumptions underlying their derivation. The study of diffusive annihilation (or 
coalescence)\footnote{Both systems belong to the same universality class \cite{Cardy} in a sense that 
the correlation functions for both systems are identical apart from the amplitude. For more details see \cite{mass2}.} is a good starting point for analyzing 
large fluctuation effects in more complicated non-equilibrium statistical systems. For example, the system $A+A \rightarrow A$ was used to analyze the aggregation
of massive diffusing particles \cite{Oleg2}.

The aim of the current paper is to study the effects of large fluctuations on correlation functions of an arbitrary order for the reactions $A+A \rightarrow \emptyset$ 
in $d\leq 2$ and $A+A+A \rightarrow \emptyset$ in $d=1$.

 Large fluctuation effects are accounted for in binary reaction-diffusion models using 
{\it Empty Interval methods} (EIM) and its generalizations \cite{mass2,mass1,ben,Doer}. This approach is restricted to $d=1$ and does not extend to higher dimensions.
There are rigorous results on the average density of particles in $d=1,2$ \cite{Bram}. The Smoluchowski approximation gives correct answers 
for average concentrations \cite{Howd} but cannot be used for higher
order correlation functions. In the early 90's, the work of Cardy and Lee \cite{Cardy,Lee} used field theoretic methods, in particular the renormalization group (RG) to obtain
an answer for the average density as well as its amplitude for $d \leq 2$. The study of $A+A \rightarrow \emptyset (A)$ has also introduced the concept of 
stochastic rate equations with {\it imaginary multiplicative noise} \cite{Cardy}.

In this paper we derive the multi-scaling of correlation functions in the systems $A+A \rightarrow \emptyset$ and $3A \rightarrow \emptyset$ using the RG method. 
The main object of our study is the large time temporal scaling of
the probability of finding $N$ particles in a small volume $\Delta V$, denoted  $P_{t}(N,\Delta V)$. 

For the $A+A \rightarrow \emptyset$ reaction-diffusion system we are 
interested in $d \leq 2$. We do not consider higher dimensions as the answers there are given by MFT. Most studies have concentrated on computing the
average density of particles ($N=1$)\cite{Cardy,Lee}.  To the best of our knowledge, the computation of multi-particle probabilities are only considered in 
\cite{mass2, mass1, ben}, with the analysis restricted to one dimension. Dynamical RG method allows us to obtain answers for the
large time limit in the form of an $\ep$-expansion ($\ep=2-d$) for $d < 2$ and logarithmic corrections to the MF scaling for $d=2$:
\begin{equation}
\label{eq:mainres1}
\frac{P_t (N, \Delta V)}{P_t (1, \Delta V)^N } \sim 	\left\{ \begin{array}{ll}
								 t^{-\frac{N(N-1)\ep}{4}+\mathcal{O}(\ep^2)}	& d < 2 \\
								(\ln t)^{-\frac{N(N-1)}{2}} \cdot \left(1+\mathcal O (\frac{1}{\ln t})\right)	& d = 2 
								\end{array}
							\right.		
\end{equation}
where \cite{Lee}
\[ P_t (1, \Delta V) \sim 	\left\{ \begin{array}{ll}
							t^{-d/2}		& d < 2 \\
							\frac{\ln t}{t} 	& d = 2 
						\end{array}
					\right.		\]
Equation (\ref{eq:mainres1}) represents the multi-scaling or the deviation of $P_t (N, \Delta V)$ from $P_t (1, \Delta V)^N$. As
\begin{equation}
\label{eq:anticor} 
\lim_{t \rightarrow \infty} \frac{P_t (N, \Delta V)}{P_t (1, \Delta V)^N} = 0 
\end{equation}
equation (\ref{eq:mainres1}) reflects the anti-correlation between particles in the large time limit. 

For the ternary reaction-diffusion system $3A \rightarrow \emptyset$ we restrict our attention to $d=1$. MFT provides the answers in higher dimensions \cite{Lee}. The
average density has been studied in \cite{Lee, Krap}, and the two-point function in \cite{Lee}. Higher order correlations have not been considered. 
The method of empty intervals 
does not extend to the ternary reaction. The RG method produces asymptotically exact results for $d=1$:
\begin{equation}
\label{eq:mainres2}
\frac{P_t (N, \Delta V)}{P_t (1, \Delta V)^N } \sim  \ln t^{-\frac{N(N-1)(N-2)}{6}}  
\end{equation}
where \cite{Lee}
\[ P_t (1, \Delta V) \sim \left[\frac{\ln t}{t}\right]^{1/2} \]

The paper is organized as follows: an introduction to the lattice model of $A+A\rightarrow \emptyset$ is given in Section II. A, its field-theoretic re-formulation is given
in Section II.B, followed by a summary of the mean field
results in Section II.C. Section III contains the RG analysis of the binary annihilation system for $d \leq 2$, starting with the description of 
renormalization procedure in Section III.A. Section
III.B contains the derivation of the corresponding Callan-Symanzik equation for the theory, its solution and the derivation of large time asymptotics of multi-particle
probabilities. We prove the exactness of our result for $P_t (N=2,\Delta V)$ 
using the first Hopf equation of the theory. In Section III.C we compare our results in $d=1$ against results from \cite{mass2} and establish their equivalence for
$N=1,2,3,4$. Further we conjecture the exactness of our one-loop answer in $d=1$ for general values of $N$. In Section IV we extend our analysis to the ternary reaction 
in $d=1$ using the same methodology. 

\section{Field-Theoretic Formulation and Mean-Field limit of $A+A \rightarrow \emptyset$ model}
\subsection{The Model}
Consider a set of point particles performing random walks, characterized 
by diffusion coefficient D, on the lattice $\mathbf{Z}^d$. Any two particles positioned at the same 
site can annihilate each other according to an exponential process with rate $\lambda$. For the simplicity of our analysis we assume finite reaction rates.
However the large time asymptotics of our model
belongs to the universality class of instantaneous annihilation-diffusion model (see end of Section III B).
It is assumed that the initial distribution of particle number at 
each site is independent Poisson with mean $N_o$. 

Let the random variable $N_t(x)$ represent the occupation number for site $x$ at time $t$.  The configuration vector 
$\underline{N}\equiv \{ N (x)\}_{x \in  \mathbf{Z}^d}$ specifies
the state of the system at time $t$ by encoding the occupation number at all sites.   $\underline{N}$ is also termed a {\it microstate}. 
Let $\mathcal{P}_t(\underline{N})$ be the probability of finding the system
in microstate $\underline{N}$ at time $t$. Correlation functions of $N_t(x)$ can be obtained by averaging functions of $N(x)$ with respect to 
$\mathcal{P}_t(\underline{N})$. For example, the average density is given by:
\begin{equation}
\label{eq:density}
\overline{N_t(x)} = \sum_{[\underline{N}]} N(x) \mathcal{P}_t (\underline{N}) 
\end{equation}
Due to translational invariance this will be independent of $x$. Our main object of interest is the large time limit asymptotics of
$\mathcal{P}_t (N(x)=N)$, the probability of finding $N$ particles at site $x$. It turns out that in the low density limit this probability is proportional to 
the $N^{th}$ {\it factorial moment} of $N_t(x)$, which we denote by $M_N(x,t)$. This can be verified as follows:
\begin{eqnarray}
\label{eq:factorial}
M_N (x,t) 	&=& \overline{[N_t(x) - (N-1)] \ldots [N_t(x) - 1] N_t(x)} \nonumber \\
	  	&=& \sum_{[\underline{N}]} \prod_{k=0}^{N-1}[N(x) - k]\cdot\mathcal{P}_{t}[\underline{N}] \nonumber \\
		&=& \sum_{[\underline{N}]}\sum_{n=0}^{\infty}\prod_{k=0}^{N-1}[n - k]\cdot \mathbf{\chi}_{[N(x)=n]}\;\mathcal{P}_{t}[\underline{N}] \nonumber \\
		&=& \sum_{n=0}^{\infty} \prod_{k=0}^{N-1}[n - k] \cdot \mathcal{P}_{t}[N(x)=n] \nonumber  \\
		&=& \sum_{n=N}^{\infty} \prod_{k=0}^{N-1}[n - k] \cdot \mathcal{P}_{t}[N(x)=n] \nonumber \\
		&\approx& N! \;\mathcal{P}_{t}[N(x) = N]
\end{eqnarray}
In the above derivation, the first five lines are exact relations. The last line is due to the fact that in the large 
time limit we expect the particle density to be low and particles are anti-correlated \cite{Cardy,mass2,ben,Lee}. Hence the configurations with the smallest possible
value of $N_t = N$ will give the dominant contribution. 

The coarse-grained counterpart of $\mathcal{P}_t [N(x)=N]$ is $P_{t}(N, \Delta V)$, the probability of finding $N$ particles in the volume element $\Delta V$. 
Let $\Delta N_t(x)$ be the number of particles in a volume $\Delta V$  (centred at $x$) at time t:
\begin{equation}
\label{eq:continuum}
\Delta N_t(x) = \int_{\Delta V} d^d y \; n_t (y)
\end{equation}
where $n_t(y)$ stands for the $density$ of particles at time $t$ at a point $y$. It should be mentioned that upon the averaging $\overline{\Delta N_t}$ is independent
of $x$ due to translational invariance. In the limit of large time and fixed $\Delta V$, 
factorial moments of $\Delta N$ are related to $P_{t}(N, \Delta V)$ via a relation analogous to (\ref{eq:factorial}):
\begin{eqnarray}
\label{eq:Nmoment}
P_{t}(N, \Delta V) &=&\frac{1}{N!}M_N (t) \nonumber \\
		 &=& \frac{1}{N!}\overline{\prod_{k=0}^{N-1}[\Delta N_t (x) - k]} 
\end{eqnarray}  

As we will demonstrate in the next section, the factorial moments of $\Delta N_{t}(x)$ admit a simple representation in terms of polynomial moments
of Doi's fields.
 
\subsection{Path-Integral representation}
In the last section it was shown that to obtain correlation functions of interest, we need to know moments of $N(x)$ with respect to $\mathcal{P}_t (\underline{N})$.
It is possible to evaluate these moments in a path-integral setting. The formalism is due to Doi-Zeldovich-Ovchinnikov and we refer the reader to \cite{Cardy} for details.
We simply present a schematic derivation. 
The time evolution of the probability measure $\mathcal{P}_t$ on the space
of microstates is given by the master equation. The master equation is a linear first order autonomous differential equation with respect to time, which implies that its
evolution operator can be expressed as a path-integral. Thus we can find a path-integral representation for any correlation function. As we are interested in universal
properties of the system at scales much larger than the lattice spacing, it is convenient to work with the continuum (coarse-grained) limit of the lattice model, 
which corresponds to some effective field theory in imaginary time. The continuum limit
in the path-integral is taken according to the rules
\[ \lambda(\Delta x)^d \rightarrow \lambda \;\;\;\;\; D(\Delta x)^2 \rightarrow D \] 
where $\Delta x$ denotes the lattice spacing. The reader is asked to refer to \cite{Lee} for more details. The resulting field theory is given
by the effective action $S$:
\begin{equation}
\label{eq:action}
S = \int_{0}^{\tau} dt \int d^{d}x \;\bar{\phi}(\partial_{t} - \Delta) \phi + 2\lambda \bar{\phi}\phi^{2} + \lambda
\bar{\phi}^{2}\phi^{2} - n_{o}\bar{\phi}\delta(\tau)
\end{equation}
where $n_o$ is the initial average particle density. We will be working in the large $n_o$ limit. The diffusion coefficient D can be set to 1 as it is not 
renormalized by fluctuations \cite{Lee}. The $\bar{\phi}$-field is a response field. The $\phi$-field is related to the local density field, $n_t(x)$ in the sense that there is
a one-to-one correspondence between correlation functions of $n_t$ and $\phi$. Let $\langle \cdot \rangle$ be the averaging with respect to the path-integral measure
$e^{-S}$. Then the average density is \cite{Lee}  
\begin{equation}
\overline{n_t(x)} = \langle\phi(x,t)\rangle\ = \int \mathcal{D}\bar{\phi}(x,t) \mathcal{D}\phi(x,t)\; \phi(x,t) e^{-S[\phi, \bar{\phi}]} 
\end{equation} 
Let us introduce the quantity $\Delta \phi$:
\begin{equation}
\label{eq:volphi}
\Delta \phi (x) = \int_{\Delta V} d^d y \;\phi (y)
\end{equation}
where the volume $\Delta V$  is centred at $x$. Averaging (\ref{eq:volphi}) results in an $x$-independent quantity due to the translational invariance of the system. 
Let us consider the statistics of $\overline{\Delta N_t(x)}$ defined in equation (\ref{eq:continuum}). Moments of $\Delta N$ and moments of $\Delta \phi$  are related as
follows
\begin{eqnarray}
\label{eq:mom1}
\langle \Delta \phi \rangle 	&=&	\overline{\Delta N_t}\\
\label{eq:mom2}
\langle \Delta \phi^2 \rangle 	&=&	\overline{\Delta N_t(\Delta N_t - 1)}
\end{eqnarray}
The relations (\ref{eq:mom1}), (\ref{eq:mom2}) follow from \cite{Lee}. Let us note that (\ref{eq:mom1}) gives us a path-integral representation of the average number of
particles in a volume $\Delta V$. More generally \cite{Oleg1} 
\begin{equation}
\label{eq:Mn}
\langle \Delta \phi^N \rangle	= M_N (t)
\end{equation}
where $M_N$ is defined by equation (\ref{eq:Nmoment}). In the previous section it was shown that
in the large-time limit the factorial moment is approximately equal to the probability of finding $N$ particles in a volume element $\Delta V$. Hence,
\begin{equation}
\label{eq:Nphi}
{P}_t (N, \Delta V) =\frac{1}{N!} \langle \Delta \phi^N  \rangle
\end{equation}
As we mentioned before, we are interested in the limit $\Delta V \rightarrow 0$.
Physically this limit corresponds to $\Delta V \ll l^d$, where the correlation length $l \sim \sqrt t$ \cite{Cardy}. By working in the large time limit we are effectively
dealing with the case $\Delta V \rightarrow 0$. Then we may write:
\begin{equation} 
\langle \Delta \phi^N \rangle =  \int_{\Delta V} dx_1\dots dx_N\; \langle\phi (x+x_1) \dots \phi (x+x_N) \rangle \\
\end{equation}
where
\begin{eqnarray}
\label{eq:wilson}
\langle \phi (x + x_1) \dots \phi (x + x_N) \rangle &=&\mathbf{F} (x_1 \dots x_N) \langle\phi^N(x)\rangle_R \cdot \nonumber \\
						& &	 (1 + \mathcal{O}(\Delta V ^{1/d}) )
\end{eqnarray}
$\mathbf{F} (x_1 \dots x_N)$ is the leading coefficient of Wilson's operator product expansion (OPE) \cite{Jean}. The limit $\Delta V \rightarrow 0$ corresponds to the 
ultraviolet ($x_{i}'s \rightarrow 0$) asymptotics of $\langle\phi (x+x_1) \dots \phi (x+x_N) \rangle$, which is given by the renormalized average of the composite
operator  $\phi^N$. We conclude that multi-scaling of $P_t (N, \Delta V)$ is described by anomalous dimensions of the composite operators $\phi^N$.
The spectrum of anomalous dimensions of $\phi^N$ will be computed in Section III.

\subsection{Mean Field Limit and Loop expansion}
\begin{figure}[h]
	\vspace{.3cm}
	\begin{center}
	\includegraphics[scale=0.64]{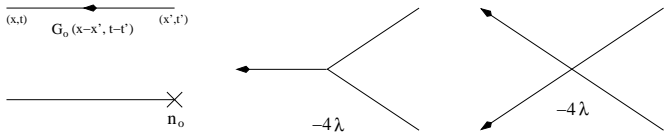}
	\end{center}
	\caption{Feynman rules for $A+A \rightarrow \emptyset$-field theory}
	\label{fig:Feyn2A}
	\end{figure}

The Feynman rules for the perturbative computation of correlation functions are derived from the action $S$, equation (\ref{eq:action}). They are given by Fig \ref{fig:Feyn2A}.
The propagator $G_o$ is the Green's function of the standard 
diffusion equation. The
perturbative expansion of $\langle \phi^m \rangle$ in powers of $\lambda$ is given by the sum of all diagrams with $m$ outgoing lines; hence
diagrams contributing to the mean density, $\langle \phi \rangle$, have one outgoing line.
The action $S$ must be dimensionless; thus in  length (L) units, the dimensions of 
the relevant parameters must be 
\begin{equation}
[t]=L^{2}\;\;\;[\bar{\phi}]=L^{0}\;\;\;[\phi]=L^{-d}\;\;\;[\lambda]=L^{d-2}.
\end{equation}
The critical dimension is $d_c = 2$, where the reaction rate is dimensionless. For $d < 2$ the field theory is {\it super-renormalizable} where all Feynman integrals
converge. For $d = 2$ we use dimensional regularization to ensure convergence of Feynman integrals.

\begin{figure}[h]
\vspace{.3cm}
\begin{center}
\includegraphics[scale=0.75]{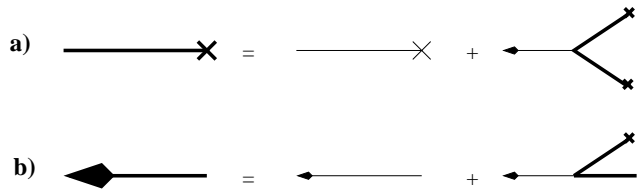}
\end{center}
\caption{a) The classical density; b) The classical response function}
\label{fig:MFeq}
\end{figure}

The bare dimensionless reaction rate is given by $g_o(t) = \lambda t^{1-d/2}$, which grows with time in $d < 2$.
A combinatorial argument shows that an $n$-loop diagram contributing to the mean density is proportional to $g_o^{n-1}(t)$ \cite{Lee}; thus in the weak coupling regime
the main contribution to the mean density comes from the sum of tree diagrams. This is equivalent to the
mean-field approximation. This sum (tree diagrams) is also termed the classical density, denoted $n_{cl}(t)$. 
In $d<2$ MFT is valid for small times given initial density is large. This agrees with our intuition: for small times local fluctuations around the large mean value of the
density are small. At large times, $g_o(t)$ grows and MFT breaks down.
To compute corrections to the mean-field answers, one must compute higher loop contributions. In particular this involves summing infinite sets 
of diagrams for a fixed number of loops. These can be re-summed in a more compact form using the classical density ($n_{cl}$)
and the classical response function ($G_{cl}$). The classical response function consists of the sum of all tree diagrams with one outgoing and one incoming line.
The diagrammatic form of the integral equations satisfied by these two quantities are given in Fig \ref{fig:MFeq}. The solution to these equations are:
\begin{eqnarray}
n_{cl}(t)	&=&	\frac{1}{2\lambda t} \\
G_{cl}(x_2,t_2;x_1,t_1)	&=& \left[\frac{n_{cl}(t_2)}{n_{cl}(t_1)}\right]^2  G_o(x_2,t_2;x_1,t_1)
\end{eqnarray}
Using $n_{cl}$ and $G_{cl}$ in combination with vertices of Fig 1, we arrive at Feynman rules generating finitely many diagrams for a given number of loops. 

We use $n_{cl}$ to write a mean-field expansion for $\langle \phi^m \rangle$ which is valid for small times in $d \leq 2$. Using dimensional analysis it can be
verified that a diagram with $m$ outgoing lines ($\langle \phi^m \rangle$) and $n$ loops is proportional to $n_{cl}^m \cdot g_o^n$. Then,
\begin{equation}
\langle \phi^m \rangle = n_{cl}^m(t) \left( 1 + \sum_{n=1}^{\infty} c_{m,n} g_o^n(t) \right)
\end{equation}
In this form we see that for small $g_o(t)$ the loop corrections are small. Then we may formulate the following mean-field answer for the probability of finding $N$
particles in volume $\Delta V$:
\begin{eqnarray}
P_{t}(N, \Delta V) &=& \frac{1}{N!}\langle \Delta \phi^N \rangle \nonumber \\
		 &\stackrel{MFT}{=}&  \frac{1}{N!}\langle \Delta \phi \rangle^N \nonumber \\
\label{eq:PNMFT}		 
		 &\sim& (\Delta V)^N t^{-N} 
\end{eqnarray}
Comparing (\ref{eq:PNMFT}) with results from \cite{mass2,mass1,ben} in $d=1$ we find the linear temporal scaling of MFT to be incorrect in the large time limit.
We will compute the correct scaling in the subsequent Section.

\section{Renormalization Group Analysis of $P_{t}(N, \Delta V)$}

The dynamical renormalization group method allows one to extract large time asymptotics of correlation functions of the theory (\ref{eq:action}). The first step is
to eliminate all $\ep \rightarrow 0$ singularities of Feynman integrals at some reference time $t_{o}$. The process of removing these divergences is called {\it renormalization}.
This is done 
by introducing a renormalized reaction rate, renormalized fields, etc. The number of renormalization constants needed to eliminate all divergences is finite
for $d\leq d_{c}$ due to renormalizability of (\ref{eq:action}). Individual terms in the renormalized perturbative series for any correlation function
$C(t)$ depend on the reference time $t_o$. The lack of 
dependence of the unrenormalized version of $C(t)$ on the unphysical parameter $t_o$ leads to renormalization group (Callan-Symanzik) equation for the correlation
function. This is solved subject to 
the initial condition at $t_o$ given by the perturbative expansion for $C(t)$ at $t=t_o$. This procedure is equivalent to re-summing all leading $\ep$-singularities
in $d<d_c$ at all orders of the loop expansion \cite{Coll}. Consequently,
one obtains scaling laws for correlation functions in terms of an $\ep$-expansion. The knowledge of $\mathcal{O}(\ep)$ terms in the loop expansion in $d<d_{c}$ yields leading
order logarithmic corrections to the mean field answers in $d=d_{c}$.

We will now apply the method described above to compute both temporal and spatial scaling exponents of $P_{t}(N, \Delta V)$ for $A+A \rightarrow \emptyset ~(d_{c}=2)$ 
reaction.
   
\subsection{One-loop renormalization of composite operator $\phi^N$}

Dimensional analysis shows that renormalization of the correlation function $\langle \prod_{i=1}^{N}\phi (x_{i},t)\rangle $, where $x_{i}\neq x_{j}$, 
requires reaction rate renormalization only \cite{Cardy,Lee}. However, we are interested in single-point correlation functions of the form $\langle \phi^N(x,t)\rangle$, see 
(\ref{eq:wilson}).
The operator $\phi^N$ is called a {\it composite operator} for $N \geq 2$. It is well known that insertion of composite operators
under the sign of averaging leads to new types of divergences in the corresponding loop expansion \cite{Jean}. These divergences cannot be eliminated
by reaction rate renormalization and require multiplicative renormalization of the corresponding composite operators. As we will see below, these extra divergences
are responsible for the multi-scaling of probabilities $P_{t}(N,\Delta V)$.   

For the theory (\ref{eq:action}) the first instance of such a divergence occurs in the correlator
$\langle\phi^2(y,t)\rangle_c = lim_{x\rightarrow y} \langle \phi(x,t)\phi(y,t)\rangle_c$ as $\ep \rightarrow 0$. We will see that it is specifically this singularity that leads to
the multi-scaling of $P_{t}(N, \Delta V)$ for $N \geq 2$. The tree-level answer is quoted below for $d<2$\cite{Lee, Mun}: 
\begin{equation}
\label{eq:ranmun}
\langle \phi(x,t)\phi(y,t)\rangle_c =\frac{1}{8 \pi g_{o}t^d}\left[\frac{1}{\varepsilon}\left[\left[\frac{|x-y|}{\sqrt t}\right]^{\varepsilon}-1\right] + 
\mathit{O}(\varepsilon^0)\right]	
\end{equation}
The averaging $\langle \cdots \rangle_c$ denotes {\it  connected} correlation functions. It is clear from equation (\ref{eq:ranmun}) that for $x \not= y$ there is 
no divergence in the limit $\varepsilon \rightarrow 0$. However if $x=y$, there will be
a $\frac{1}{\ep}$ divergence (as $\ep \rightarrow 0$) despite (\ref{eq:ranmun}) being a tree-level answer. This divergence cannot be regularized by reaction rate renormalization.
It requires {\it multiplicative renormalization}, where the divergent function 
is multiplied by a renormalizing factor which will eliminate the singularity. 
This procedure violates naive dimensional arguments according to which $\langle \phi ^N \rangle$ should scale as $\langle \phi \rangle \sim t^{-Nd/2}$ for $d < 2$. 
Multiplicative renormalization leads to a non-trivial anomalous dimension which 
depends non-linearly on the order of the composite operator. Multiplicative renormalization  
is not required for the average density $\langle \phi \rangle$ which explains its lack of anomalous scaling \cite{Cardy}. 

\begin{figure}[h!]
\vspace{.3cm}
\begin{center}
\includegraphics[scale=0.5]{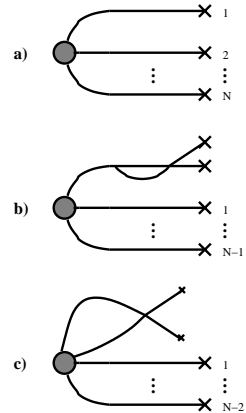}
\end{center}
\caption{The diagrams contributing to $\langle \phi^N \rangle$ to $\mathcal{O}(g^{1-N})$}
\label{fig:2AN}
\end{figure}

Let $M_N(t)=\langle\phi^N(t)\rangle$. 
Our aim is to renormalize $M_N$ at (arbitrary reference) time $t_o$. Let $g_o = \lambda t_o^{\varepsilon/2}$ be the
bare (dimensionless) reaction rate. The renormalized rate $g$ is also defined at reference time $t_o$ and is given by \cite{Lee}:
\begin{equation}
\label{eq:recoup}
g = \frac{g_o}{1 + g_o/g^*}
\end{equation}
where $g^* =  \frac{\ep}{C_d} = 2\pi\varepsilon + \mathcal{O}(\varepsilon^2)$ is the non-trivial stable fixed point of the RG flow in the space of effective reaction rates.  
The constant $C_d$ is given by
\begin{equation}
\label{eq:cd}
C_d		= \frac{2 \ep}{(8 \pi)^{d/2}}\Gamma \left( \frac{\ep}{2} \right)
\end{equation}
$C_d$ is regular at $d=2$ and takes the value $\frac{1}{2 \pi}$.
Expressing $g_o$ as a power series in $g$ and substituting into $M_N(t_o)$, enables us to obtain $M_N$ as a power series in $g$:
\begin{equation}
\label{eq:pertexp}
M_N(t_o) =\left[n_{cl}(g_o, t_o)\left.\right|_{g_o=g}\right]^N \left[ 1 + \sum_{n=1}^{\infty} c_{N,n} g^n \right]
\end{equation}
The expression for $M_N(t_o)$ to $\mathcal{O}(g^{1-N})$ is summarized by the 
diagrams in Fig. \ref{fig:2AN}. 
The first term in this series is just $n_{cl}^N$ and is given by Fig \ref{fig:2AN}a.  
Expanding $n_{cl}$ in $g$ using relation (\ref{eq:recoup}) will generate a term proportional to $\frac{1}{g^*} \sim \frac{1}{\ep}$. This term acts
as a counter term to the diagram shown in Fig \ref{fig:2AN}b. It eliminates the singularity in the one-loop correction to mean density $n^{(1)}$ \cite{Lee}. 
The first divergence which cannot be eliminated by the renormalized reaction rate, appears in the coefficient $c_{N,1}$. The origin of this
singularity is from averaging the composite operator $\phi^2$. Let $Z_N$ be the renormalizing factor of $M_N$. 
\begin{equation}
Z_N = 1 + \sum_{n=1}^{\infty} a_{N,n} g^n
\end{equation}
It is chosen in such a way that the renormalized correlation function $M_{N,R}$ is non-singular at $\ep =0$ at all orders. By definition
\footnote{Lack of operator mixing produces this simple form for equation (\ref{eq:phiNre}). This is due to the fact that in our theory
all processes (diagrams) reduce the particle number. For more see \cite{Oleg1}.}
\begin{equation}
\label{eq:phiNre}
M_{N,R}(g, t_o)	=  Z_N(g,\varepsilon) \cdot M(g,t_o, \ep) 
\end{equation}
The coefficients $a_{N,n}$ are chosen to cancel the singularities in the coefficients $c_{N,n}$. This is known as the {\it minimal subtraction} scheme. In particular
\begin{equation}
\label{eq:singu}
a_{N,1} = -\left[c_{N,1}\right]_s
\end{equation}
where $[\cdot]_s$ extracts the singular part of an expression. Let us look at the calculation in more detail. Using Fig. (\ref{fig:2AN}) we may write
\begin{equation}
M_N (t_o) = \left[\frac{1}{2gt_o^{d/2}}\right]^N \left[ 1 - \frac{N(N-1)}{4 \pi \ep}\cdot g + \mathcal{O}(g^2) + \mathrm{finite} \right]
\end{equation}
The $\mathcal{O}(g)$ term is computed as follows: we need $n_{cl}^{(N-2)}$, there are $_N C_2$ diagrams of type c and finally by setting $x = y$ in (\ref{eq:ranmun})
which gives $[M_2(t_o)]_s = \frac{-1}{8\pi\varepsilon}\;t_o^{-d}$. Hence
\begin{equation}
\label{eq:zn}
Z_N = 1 + \frac{N(N-1)}{4\pi\varepsilon} \cdot g + \mathcal{O}(g^2)
\end{equation}

\subsection{RG computation of Multi-Scaling} 
The knowledge of renormalization laws (\ref{eq:recoup}) and (\ref{eq:zn}) can be used to compute $M_{N,R}(t)$ for $t > t_o$ as follows. 
The Markov property of the evolution operator $U$ for the bare theory, namely $U(t,t_o)U(t_o,0)=U(t,0)$, tells us that the bare function
$M_N(t)$ is independent of $t_o$ for $t > t_o$. Hence
\begin{equation}
\label{eq:CSpde1}
t_o\frac{\partial}{\partial t_o}M(t) = t_o\frac{\partial}{\partial t_o}\left[ Z_N^{-1}\cdot M_{N,R}(t,t_o) \right] = 0 
\end{equation}
The function $M_{N,R}(t)$ is a function of $(t,t_o, g)$, leading to the ansatz: $M_{N,R}(t) = t_o^{-Nd/2}\Phi\left(\frac{t}{t_o},g(t_o)\right)$, where $\Phi$ is a
function with dimensionless arguments. The choice of $t_o$ is arbitrary, but we choose it small enough so that MFT is still valid. This motivates the
choice of the pre-factor $t_o^{-Nd/2}$.
Upon substitution of this ansatz into (\ref{eq:CSpde1}) we obtain the Callan-Symanzik equation:
\begin{equation}
\label{eq:cs}
\left[ t\frac{\partial}{\partial t} + \beta (g) \frac{\partial}{\partial g} + \frac{Nd}{2} + \gamma_N (g) \right] M_{N,R}(t,t_o,g) = 0
\end{equation}
where the $\beta$ and $\gamma_N$ functions of the theory are given by
\begin{eqnarray}
\beta(g) 	&=& -t_o \frac{\partial g}{\partial t_o} = \frac{1}{2}\left[C_d g^2-\ep g \right] \\
\gamma_N(g) 	&=& -\beta(g)\frac{\partial \ln Z_N}{\partial g} = \frac{N(N-1)}{8 \pi} g + \mathcal{O}(g^2)
\end{eqnarray}
The initial condition (at time $t_o$) is given by the loop expansion of $\langle \phi^N \rangle$ with the most dominant contribution coming from the MFT answer:
\begin{equation}
\label{eq:initl}
M_{N,R} (t_o, g) = n_{cl}^N(t_o, g)
\end{equation}
Equation (\ref{eq:cs}) subject to initial condition (\ref{eq:initl}) is solved using the
method of characteristics and has the following solution for $d<2$:
\begin{eqnarray}
\label{eq:solnu2}
M_{N,R}(t,g(t,t_o)) &=& \left(\frac{t_o}{t}\right)^{Nd/2} n^N(t_o, g(t,t_o)) \cdot  \nonumber \\
			& &
\left[\frac{g(t,t_o)-g^*}{g-g^*}\right]^{\frac{N(N-1)}{4 \pi\ep}g^*}\\
\label{eq:runu2}
g(t,t_o) &=& \frac{g^*}{1-  \left(1-\frac{g^*}{g}\right) \left(\frac{t_o}{t}\right)^{\varepsilon/2} } 
\end{eqnarray}
The {\it running coupling} $g(t,t_o)$ is the effective RG flow of the reaction rate. 
It can be verified that  $\lim_{t \rightarrow \infty} g(t,t_{o}) = g^*$, which is of order $\ep$. 
Hence for large times we can convert the loop expansion 
to an $\varepsilon$-expansion.
To obtain answers in $d=2$ we 
take the limit $\ep \rightarrow 0$ in equations (\ref{eq:solnu2}), (\ref{eq:runu2}):
\begin{eqnarray}
M_{N,R}(t,g(t,t_o)) &=& \left(\frac{t_o}{t}\right)^{N} n^N(t_o, g(t,t_o)) \cdot \nonumber \\
			& & \left[\frac{g(t,t_o)}{g}\right]^{\frac{N(N-1)}{2}} \\
g(t,t_o) &=& \frac{g}{1+\frac{g}{4 \pi} \ln \left(\frac{t}{t_o}\right)}
\end{eqnarray}
For large times $g(t,t_o) \sim \frac{4 \pi}{\ln \left(\frac{t}{t_o}\right)}$. Then in the large time limit we obtain the following scaling in $t$ 
(recall $M_{N,R}(t) \equiv \langle \phi^N(t) \rangle_R$) :
\begin{equation}
\label{eq:MN}
M_{N,R}(t)			\sim \left\{ \begin{array}{ll}
					t^{-Nd/2}t^{-\frac{N(N-1)\ep}{4}+\mathcal{O}(\ep^2)} 	& d < 2 \\
					 \left(\frac{\ln t}{t}\right)^N \left(\ln t \right)^{-\frac{N(N-1)}{2}}\cdot \left(1+\mathcal O (\frac{1}{\ln t})\right) & d = 2 
					\end{array}
					\right.		
\end{equation}
Combining (\ref{eq:Nphi}) and (\ref{eq:MN}) we obtain the following results for $P_t(N,\Delta V)$  
\begin{equation}
\label{eq:PN}
\frac{P_t(N,\Delta V)}{P_t(1,\Delta V)^N}
	\sim \left\{ \begin{array}{ll}
					\left( \frac{\Delta V^{2/d}}{t}\right) ^{\frac{N(N-1)\ep}{4}+\mathcal{O}(\ep^2)} 	& d < 2 \\
					 \left( \ln \left[\frac{t}{\Delta V}\right] \right)^{\frac{-N(N-1)}{2}}\cdot \left(1+\mathcal O (\frac{1}{\ln t})\right) 	& d = 2 
					\end{array}
					\right.		
\end{equation}
where the $\Delta V$ dependence is restored using dimensional arguments.
The physical interpretation of (\ref{eq:PN}) is that in the large-time limit particles are {\it anti-correlated} (recall equation (\ref{eq:anticor})):
given the same average density the probability of finding $N$ reacting particles
in $\Delta V$ goes to zero faster than the probability of finding $N$ non-reacting particles in $\Delta V$. The origin of anti-correlation can be traced back
to the recurrence property of random walks in $d \leq 2$.

Let us compare (\ref{eq:PN}) with exact results for $N=1,2$.
For $N=1$ there is no anomaly and our formula is in agreement with the well-known result found in \cite{Lee} for 
$d \leq 2$. Let us examine the case $N=2$ by studying the Langevin SDE : $\partial_t \phi = -2 \lambda \phi^2$ + Ito noise term \cite{Oleg2}, which is equivalent
to the field theory (\ref{eq:action}). 
Taking averages of both sides yields 
the first Hopf equation: $\partial_t \langle\phi\rangle = -2 \lambda \langle\phi^2\rangle$. For $d < 2$ we know $\langle\phi\rangle \sim t^{-d/2}$. Substituting this into the 
Hopf equation gives $\langle\phi^2\rangle \sim t^{-(1+d/2)}=t^{-(d+\varepsilon/2)}$. In $d=2$, $\langle\phi\rangle 
\sim \frac{\ln t}{t}$ and using this in the Hopf equation yields $\langle\phi^2\rangle \sim \frac{\ln t}{t^2}$. These are exact relations and are 
in agreement with our formula (\ref{eq:MN}) with $\mathcal{O}(\varepsilon ^2) = 0$.
We conjecture that for $N=2$, the $\mathcal{O}(\ep^2)$ corrections are absent in (\ref{eq:MN}).
Finally, note from equation
(\ref{eq:recoup}) $\lim_{\lambda \rightarrow \infty} g = g^*$. Thus for $d \leq 2$, the limit $\lambda \rightarrow \infty$ for finite $t$ yields the same asymptotics as 
the limit $t \rightarrow \infty$ for finite $\lambda$. Therefore the large time asymptotics of the model at hand belongs to the universality class of instantaneous annihilation. 

\subsection{Comparison of results of $\ep$-expansion with exact results in d=1}
Equation (\ref{eq:PN}) gives us an asymptotically exact result in $d=2$ $(\ep = 0)$. In the previous section we have also confirmed that order-$\ep$ expansion of scaling
exponent of $P(2, \Delta V)$ given by the first equation in (\ref{eq:PN}) is exact in all dimensions $d \leq 2$. As it turns out, (\ref{eq:PN}) yields an exact answer
for multi-scaling of probabilities in $d=1$ for all values of $N$.    
We are fortunate to have some exact results for multi-point correlation functions in the problem of diffusion-limited annihilation $A+A \rightarrow \emptyset$ for $d=1$,
$\lambda = \infty$ \cite{mass2}.
For more details please refer to \cite{mass2,mass1,ben}. Using Mathematica and recurrence relations derived in \cite{mass2} we were able to  
compute exact analytical expressions for $P_t(N,\Delta V)$ for $N=1,2,3,4$. Based on the results we find that the 
$\mathcal{O}(\ep^2)$ terms are absent in
the $\ep$-expansion (\ref{eq:PN}). This leads us to conjecture that
in $d=1$ the one-loop answer for the scaling exponents are exact. 

To substantiate our claims, let us review the key results from \cite{mass2}.
The reaction rate is infinite, hence we do not expect to find more than one particle at a given site. We will use the notation used in
\cite{mass2,mass1,ben}. The correlation function $\rho_N(x_1, \ldots ,x_N; t)$ 
represents the joint probability density of finding N particles positioned at $x_1, \ldots ,x_N$ at time $t$. In particular, $\rho(t) \equiv \rho_1(t)$ is the average density. 
There is also the convention that $x_i < x_j$ for $i < j$. 
In the limit of large times:
\begin{equation}
\label{eq:pconvP}
P_t(N,\Delta V) = \int_{\Delta V} dx_1\ldots dx_N \; \rho_N(x_1, \ldots ,x_N; t)
\end{equation}
In the large time limit the following answers hold true:
\begin{eqnarray}
\rho(t)						&=&	\frac{1}{\sqrt{8\pi Dt}} \\
\frac{\rho_2(x_1, x_2; t)}{\rho^2(t)}		&=&	1 - e^{-2z_{21}^2} + \sqrt{\pi}z_{21} e^{-z_{21}^2}\mathrm{erfc}(z_{21}) \\
z_{ji}						&=& 	\frac{x_j - x_i}{\sqrt{8Dt}},
\end{eqnarray}
where $D$ is the diffusion coefficient \cite{mass2}.
Note that as $x_2 \rightarrow x_1$ (or vice-versa), the correlation function vanishes. This is a reflection of anti-correlations between particles. 
For small separations, it can be easily shown that 
\begin{equation}
\label{eq:M2}
\frac{\rho_2(x_1, x_2; t)}{\rho^2(t)}	=  \sqrt{\pi}z_{21} + \mathcal{O}(z_{21}^2)
\end{equation}
The above expression for $\rho_2$ is valid in the limit of large time and fixed separation $\Delta=x_2-x_1$, as in this limit $z_{21}\rightarrow 0$. 
Due to (\ref{eq:pconvP}) the scaling for (\ref{eq:M2}) agrees with our answer for $N=2$ in (\ref{eq:PN}):
\begin{equation}
\frac{P_t(2,\Delta V)}{P_t(1,\Delta V)^2} \sim  \frac{\Delta}{t^{1/2}}
\end{equation}
With the aid of Mathematica and using similar arguments as above, it can be shown that $\frac{\rho_3(x_1,x_2,x_3;t)}{\rho^3(t)} \sim t^{-3/2}$ and 
$\frac{\rho_4(x_1,x_2,x_3,x_4;t)}{\rho^4(t)} \sim t^{-3}$ , which are in agreement with our conjecture for the cases $N=3$ and $N=4$ respectively.

The formal reason for anomalous scaling for $N=2$ is vanishing of the two-particle probability distribution function at $x_1=x_2$. This is the most
clear indication of anti-correlation between annihilating particles. This same phenomenon is responsible for zeros of multi-particle distribution functions as well.
To explore the nature of these zeros starting from the exact recurrence relations of \cite{mass2} we were forced to use Mathematica:
exact expressions for correlation functions are written as linear combination of a variety of terms involving
products of $\mathrm{exp} (-z^2_{ji}), \mathrm{erfc} (z_{ji}), z_{ji}$. At small separations these expressions simplify due to a number of cancellations
which are not obvious. Using Mathematica to break through the tedious computations we find:
\begin{eqnarray}
\frac{\rho_2(x_1, x_2; t)}{\rho^2(t)}	&=&  \sqrt{\pi}z_{21} \\
\frac{\rho_3(x_1,x_2,x_3;t)}{\rho^3(t)} &=&  6\sqrt{\pi}z_{21}z_{31}z_{32} \\
\frac{\rho_4(x_1,x_2,x_3,x_4;t)}{\rho^4(t)} &=&  2\pi z_{21}z_{31}z_{32}z_{41}z_{42}z_{43}\label{n234}
\end{eqnarray}
 
Note that all distribution functions above vanish as the first power of separation between any pair of particles.
Hence we can make a conjecture about spatio-temporal behaviour of distribution functions for arbitrary $N$:
\begin{equation}
\label{eq:conj}
\frac{\rho_N(x_1,...,x_N;t)}{\rho^N(t)}	\sim \prod_{1 \leq i < j \leq N} z_{ji} 
\end{equation}
The above expression is a direct generalization of (\ref{n234}) based on permutation symmetry and self-similarity. It states that the spatial dependence 
of the probability density is given by the Van-der-Monde determinant of particles' coordinates.

Conjecture (\ref{eq:conj}) reproduces the temporal scaling derived in (\ref{eq:MN}) as the right hand side contains $N(N-1)/2$ factors
 $z_{ji}$ each  contributing $t^{-1/2}$ to the scaling. We will present a rigorous proof of this conjecture in a separate publication.

\section{Reaction $3A \rightarrow \emptyset$ in critical dimension $d=1$}

\begin{figure}[h]
\vspace{.3cm}
\begin{center}
\includegraphics[scale=0.75]{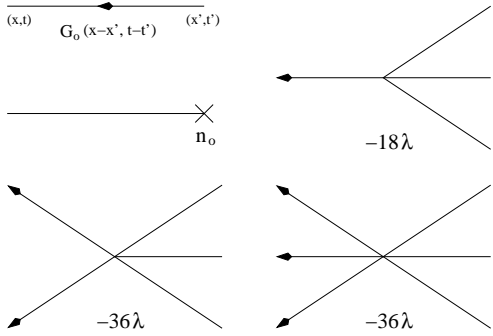}
\end{center}
\caption{Feynman rules for $3A \rightarrow \emptyset$-field theory}
\label{fig:Feyn3A}
\end{figure}

In the final section of this paper, we extend the preceding analysis to the problem $3A \rightarrow \emptyset$, where reactions occur in triples.
Unlike the binary case we cannot apply EIM to analyze this reaction in $d=1$. Thus we will adopt the field-theoretic approach. We give a
brief introduction to the important quantities in the model \cite{Lee}. The action $S$ given by 
\begin{eqnarray}
S &=& \int d^dx \int_{0}^{\tau} dt \; \bar{\phi}(\partial_{t} - \Delta)\phi + 3\lambda \bar{\phi} \phi^3 + 3 \lambda \bar{\phi}^2 \phi^3 + \nonumber \\
& & \lambda \bar{\phi}^3 \phi^3 - n_0 \bar{\phi} \delta(t)
\end{eqnarray}
The Feynman rules are shown in Fig \ref{fig:Feyn3A}. The action must be dimensionless which requires the following:
\begin{equation}
[t]=L^2\;\;\;[\phi]=L^{-d}\;\;\;[\bar{\phi}]=L^{0}\;\;\;[\lambda]=L^{2d-2}
\end{equation}

\begin{figure}[h]
\vspace{.3cm}
\begin{center}
\includegraphics[scale=0.75]{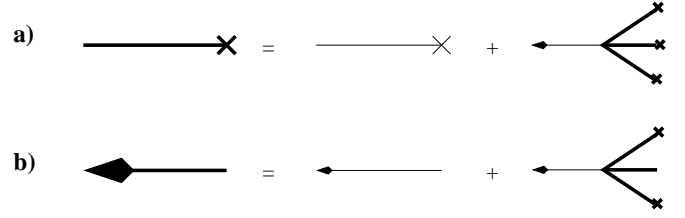}
\end{center}
\caption{a) The classical density; b) The classical response function}
\label{fig:MFeq3}
\end{figure}

The reaction rate is dimensionless at $d=1$, which is the critical dimension for this problem. Hence in $d=1$ we expect the system to be characterized by
MFT with logarithmic corrections. As for the binary system, we use the classical (MF) versions of the density and response functions to study the loop expansion. 
The integral equations satisfied by the
classical density and classical response function are shown in Fig \ref{fig:MFeq3}. Their solutions are given by
\begin{eqnarray}
n_{cl}(t)	&=&	\left[\frac{1}{6\lambda t}\right]^{1/2} \\
G_{cl}(x_2,t_2;x_1,t_1)	&=& \left[\frac{n_{cl}(t_2)}{n_{cl}(t_1)}\right]^3  G_o(x_2,t_2;x_1,t_1)
\end{eqnarray}

Power counting shows that reaction rate renormalization takes care of singularities of Greens' functions which do not contain any composite operators.
Moreover it turns out that all
divergences of $\langle \phi^2 \rangle$ are also eliminated by reaction rate renormalization. For higher order composite operators
one needs multiplicative renormalization. 

\begin{figure}[h]
\vspace{.3cm}
\begin{center}
\includegraphics[scale=0.5]{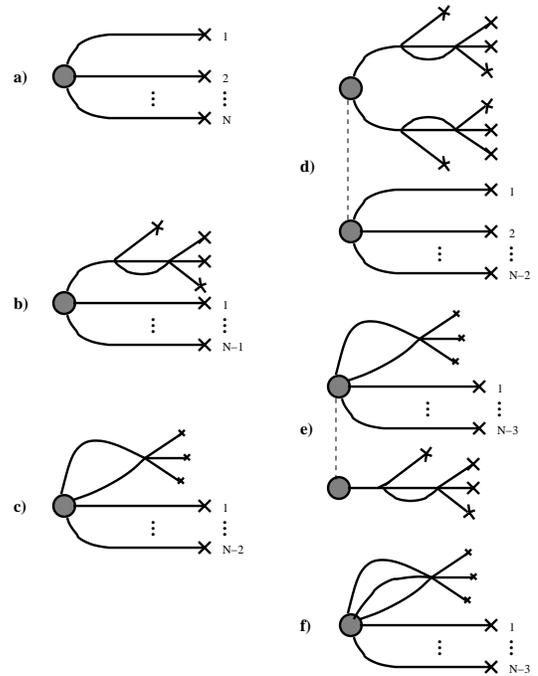}
\end{center}
\caption{The diagrams contributing to $\langle \phi^N \rangle$ to $\mathcal{O}(g^{1-N/2})$}
\label{fig:3AN}
\end{figure}

Let $g_o = \lambda t_o^{\varepsilon}$ be the bare (dimensionless) reaction rate, where $t_o$ is our arbitrary reference time.
Note that here $\ep = 1 -d$. To eliminate divergences associated with reaction rate renormalization we need an expression for $g_o$ in terms of $g$.
The relation (\ref{eq:recoup}) holds, 
but $g^* = \frac{2 \pi \varepsilon}{\sqrt3} + \mathcal{O}(\varepsilon^2)$.
The lowest order diagrams contributing to $M_N \equiv \langle \phi^N \rangle$ are shown in Fig \ref{fig:3AN}. The only divergence at the 1-loop level 
associated with composite
operators comes from the connected three-point
function. This determines $Z_N$ and hence the anomalous scaling.
The Callan-Symanzik equation in $d=1$ takes the form
\begin{equation}
\label{eq:cs3A}
\left[ t\frac{\partial}{\partial t} + \beta (g) \frac{\partial}{\partial g} + \frac{N}{2} + \gamma_N (g) \right] M_{N,R}(t,t_o,g) = 0
\end{equation}
where
\begin{eqnarray}
\beta(g) 	&=& \frac{\sqrt3}{2\pi}g^2 \\
\gamma_N(g) 	&=& \frac{N(N-1)(N-2)}{4 \pi\sqrt3} g + \mathcal{O}(g^2)
\end{eqnarray}
Once again we choose the reference time $t_o$ small so that the dominant contribution at $t_o$ is given by MFT
\begin{equation}
M_{N,R} (t_o, g) = n_{cl}^N (t_o, g)
\end{equation}
The solution of (\ref{eq:cs3A}) is given by
\begin{eqnarray}
M_{N,R}(t,g(t,t_o)) &=& \left(\frac{t_o}{t}\right)^{N/2} n_{cl}^N(t_o, g(t,t_o)) \cdot \nonumber \\
			& & \left[\frac{g(t,t_o)}{g}\right]^{\frac{N(N-1)(N-2)}{6}} \\
g(t,t_o) &=& \frac{g}{1+\frac{\sqrt3 g}{2 \pi} \ln \left(\frac{t}{t_o}\right)} 
\end{eqnarray}
Recall $M_{N,R} (t) \equiv  \langle \phi^N(t) \rangle_R$. For large times, we obtain the following scaling behaviour of composite operators:
\begin{eqnarray}
\label{eq:anom}
\langle \phi^N(t) \rangle_R &\sim& A_N \left[\frac{\ln t}{t}\right]^{N/2} \cdot (\ln t)^{\frac{-N(N-1)(N-2)}{6}}\cdot \nonumber \\ 
						& & \left(1+\mathcal O \left(\frac{1}{\sqrt{\ln t}}\right)\right) 
\end{eqnarray}
where
\begin{equation}
A_N = \left(\frac{1}{\sqrt{6}}\right)^{N}\cdot \left(\frac{\sqrt 3}{2 \pi}\right)^{\frac{3N-N(N-1)(N-2)}{6}} 
\end{equation}
Using the relation (\ref{eq:Nphi}) between composite operators and probabilities
\begin{equation}
\label{eq:PN3}
\frac{P_t(N,\Delta V)}{P_t(1,\Delta V)^N} \sim \left(\ln \left[\frac{t}{\Delta V}\right] \right)^{\frac{-N(N-1)(N-2)}{6}}\cdot 
\left(1+\mathcal O \left(\frac{1}{\sqrt{\ln t}}\right)\right) 
\end{equation}
The result (\ref{eq:PN3}) reflects the anti-correlation of particles as stated in equation (\ref{eq:anticor}). 
We know from \cite{Lee} that $\langle \phi \rangle \sim \left[\frac{\ln t}{t}\right]^{1/2}$ which agrees with (\ref{eq:anom}) 
upon setting $N=1$. There is no anti-correlation between pairs of particles as
reactions only occur in triples. Therefore the anomaly should vanish for $N=2$ which agrees with (\ref{eq:anom}),(\ref{eq:PN3}).
Due to the absence of singularities associated with $\langle \phi^2 \rangle$ we know the dominant contribution comes from the disconnected $n_{cl}$ diagrams for $N=2$. 
Then $\langle \phi^2 \rangle \sim \frac{\ln t}{t}$ \cite{Lee}
which is also in agreement with (\ref{eq:anom}). Finally to check (\ref{eq:anom}) for $N=3$, we can use the first Hopf equation for the theory: 
$\partial_t \langle \phi \rangle = -3\lambda\langle \phi^3 \rangle$. As $\langle \phi\rangle \sim \left[\frac{\ln t}{t}\right]^{1/2}$, the Hopf
equation implies that $\langle \phi^3 \rangle \sim t^{-3/2}(\ln t)^{1/2}$ exactly. This agrees with (\ref{eq:anom}) as well.
\section{Summary}
The main finding of this paper was the multi-scaling of the probability distributions of multi-particle configurations for single species reaction-diffusion 
systems. The scaling was indicative of particles being anti-correlated in the large time-limit. In particular, the quadratic scaling exponent in the binary system
reflects pairwise anti-correlation. For the ternary case, the scaling exponent is cubic which shows anti-correlation within particle triples.
 We obtained our results in a field-theoretic setting by identifying
probability distributions of multi-particle configurations, at scales much smaller than correlation length, with composite operators in Doi-Zeldovich field theory.
The origin of the multi-scaling can therefore be traced back to the anomalous dimensions of the corresponding composite operators.

We obtained exact logarithmic corrections to scaling  for the binary system in $d=2$ and for the ternary reaction in $d=1$. We computed scaling exponents for the binary system
in $d< 2$ using $\ep$-expansion. By analyzing the first Hopf equation for the binary system, we proved that the one-loop $\ep$-expansion gives the exact answer
for the probability of finding two particles in a fixed volume. A similar computation for the ternary system confirms the result of the RG computation for the probability 
of finding three particles in the fixed volume.

RG analysis led us to several conjectures
for scaling exponents for the $A+A \rightarrow \emptyset$ system in $d=1$. First, by comparing one-loop answers with the exact results in one
dimension \cite{mass2} for $N=1,2,3,4$, we conjecture that two and higher loop corrections are absent in $d=1$ for an arbitrary number of particles $N$, see (\ref{eq:PN}).
Second, basing on the previous conjecture and dimensional analysis we propose that the spatial dependence of the multi-particle probability density
is given by the absolute value of the Van-der-Monde determinant of particles' positions, see (\ref{eq:conj}). We have recently found a rigorous proof of the stated conjectures,
which will be published separately.

\section{Acknowledgments}
 We would like to thank Colm Connaughton for useful discussions and help with Mathematica, and Roger
Tribe for numerous useful discussions. R. M.  gratefully acknowledges MIND for support of this research.
  
\pagebreak

\end{document}